\documentclass{ieeetran}

\usepackage{microtype}
\usepackage{graphicx}
\usepackage{subfigure}
\usepackage{booktabs} % for professional tables

\usepackage{hyperref}

\usepackage{color} % test only

\usepackage{xspace}
\newcommand{\PPGINI}{\ensuremath{\pi_{\mathsf{MS-GINI}}}\xspace}
\newcommand{\PPGINIFS}{\ensuremath{\pi_{\mathsf{GINI-FS}}}\xspace}
\newcommand{\PPFILTERFS}{\ensuremath{\pi_{\mathsf{FILTER-FS}}}\xspace}
\newcommand{\SLT}{S_{\leq \theta}}
\newcommand{\SGT}{S_{> \theta}}
\newcommand{\peq}{\ensuremath{\pi_{\mathsf{EQ}}}\xspace}
\newcommand{\pDP}{\ensuremath{\pi_{\mathsf{DP}}}\xspace}
\newcommand{\pDIV}{\ensuremath{\pi_{\mathsf{DIV}}}\xspace}
\newcommand{\plt}{\ensuremath{\pi_{\mathsf{LT}}}\xspace}
\newcommand{\pmul}{\ensuremath{\pi_{\mathsf{DM}}}\xspace}
\newcommand{\pmmul}{\ensuremath{\pi_{\mathsf{DMM}}}\xspace}

\newcommand{\pargmin}{\ensuremath{\pi_{\mathsf{ARGMIN}}}\xspace}

\def\baselinestretch{0.98}

\usepackage{algorithm}
\usepackage{algorithmic}
\newenvironment{protocol}[1][]{%
    \floatname{algorithm}{Protocol}% Update algorithm name
   \begin{algorithm}[#1]%
   \footnotesize{}
  }{\end{algorithm}}
  
\usepackage{amsthm}
\theoremstyle{definition}

\usepackage{enumitem}

\usepackage{amsthm,amsmath,amssymb}
\usepackage{mathrsfs}
\usepackage{soul}
\begin{document}

\title{Privacy-Preserving Feature Selection with Secure Multiparty Computation}

\author{Xiling Li, Rafael Dowsley and Martine De Cock
\thanks{Xiling Li is with the School of Engineering and Technology, University of Washington, Tacoma, WA, USA. Email: xl32@uw.edu}
\thanks{Rafael Dowsley is with the Faculty of Information Technology, Monash University, Clayton, Australia. Email: rafael.dowsley@monash.edu}
\thanks{Martine De Cock is with the School of Engineering and Technology, University of Washington, Tacoma, WA, USA and Ghent University, Ghent, Belgium. Email: mdecock@uw.edu}
}

\maketitle

\begin{abstract}
Existing work on privacy-preserving machine learning with Secure Multiparty Computation (MPC) is almost exclusively focused on model training and on inference with trained models, thereby overlooking the important data pre-processing stage. In this work, we propose the first MPC based protocol for private feature selection based on the filter method, which is independent of model training, and can be used in combination with any MPC protocol to rank features. We propose an efficient feature scoring protocol based on Gini impurity to this end. To demonstrate the feasibility of our approach for practical data science, we perform experiments with the proposed MPC protocols for feature selection in a commonly used machine-learning-as-a-service configuration where computations are outsourced to multiple servers, with semi-honest and with malicious adversaries.
Regarding effectiveness, we show that secure feature selection with the proposed protocols improves the accuracy of classifiers on a variety of real-world data sets, without leaking information about the feature values or even which features were selected. Regarding efficiency, we document runtimes ranging from several seconds to an hour for our protocols to finish, depending on the size of the data set and the security settings.
\end{abstract}

\IEEEpeerreviewmaketitle

\section{Introduction}
Machine learning (ML) thrives because of the availability of an abundant amount of data, and of computational resources and devices to collect and process such data. In many effective ML applications, the data that is consumed during ML model training and inference is often of a very personal nature. Protection of user data has become a significant concern in ML model development and deployment, giving rise to laws 
to safeguard the privacy of users, such as the European General Data Protection Regulation (GDPR)
and the California Customer Privacy Act (CCPA).
Cryptographic protocols that allow computations on encrypted data are an increasingly important mechanism to enable data science applications while complying with privacy regulations.
In this paper, we contribute to the field of privacy-preserving machine learning (PPML), a burgeoning and interdisciplinary research area at the intersection of cryptography and ML that has gained significant traction in tackling privacy issues.

\begin{figure*}
  \centering
  \includegraphics[width=\linewidth]{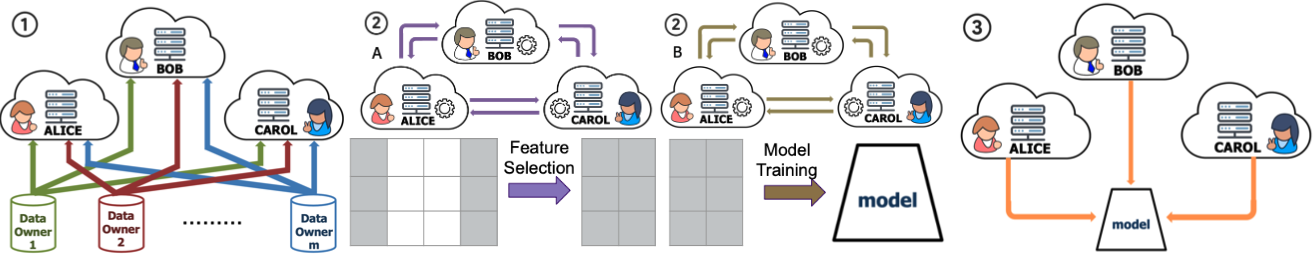}
  \caption{Overview of private feature selection and model training in 3PC setting with computing servers (parties) \textit{Alice}, \textit{Bob}, and \textit{Carol}.}
  \label{fig:3pc}
\end{figure*}

In particular, we use techniques from Secure Multiparty Computation (MPC), an umbrella term for cryptographic approaches that allow two or more parties to jointly compute a specified output from their private information in a distributed fashion, without actually revealing their private information to each other \cite{mpc_book}. We consider the scenario where different data owners or enterprises are interested in training an ML model over their combined data. There is a lot of potential in training ML models over the aggregated data from multiple enterprises. First of all, training on more data typically yields higher quality ML models. For instance, one could train a more accurate model to predict the length of hospital stay of COVID-19 patients when combining data from multiple clinics. This is an application where the data is \textit{horizontally distributed}, meaning that each data owner or enterprise has records/rows of the data. Furthermore, being able to combine different data sets enables new applications that pool together data from multiple enterprises, or even from different entities within the same enterprise. An example of this would be an ML model that relies on lab test results as well as healthcare bill payment information about patients, which are usually managed by different departments within a hospital system. This is an example of an application where the data is \textit{vertically distributed}, i.e.~each data owner has their own columns.
While there are clear advantages to training ML models over data that is distributed across multiple data owners, often these data owners do not want to disclose their data to each other, because the data in itself constitutes a competitive advantage, or because the data owners need to comply with data privacy regulations. These roadblocks can even affect different departments within the same enterprise, such as different clinics within a healthcare system.

During the last decade, cryptographic protocols designed with MPC have been developed for training of ML models over aggregated data, without the need for the individual data owners or enterprises to reveal their data to anyone in an unencrypted manner. This existing work includes MPC protocols for training of 
decision tree models \cite{lindell2000privacy,de2014practical,Choudhary:2020,escudero2020}, 
linear regression models \cite{nikolaenko2013privacy,AISec:CDNN15,agarwal2019protecting},
and neural network architectures 
\cite{7958569,agrawal2019quotient,wagh2019securenn,guo2020secure,BMCMG:CDNRST21}.
Existing approaches assume that the data sets are pre-processed and clean, with features that have been pre-selected and constructed. In practical data science projects, model building constitutes only a small part of the workflow: real-world data sets must be cleaned and pre-processed, outliers must be removed, training features must be selected, and missing values need to be addressed before model training can begin. Data scientists are estimated to spend 50\% to 80\% of their time on data wrangling as opposed to model training itself \cite{NYT:2014}.
PPML solutions will not be adopted in practice if they do not encompass these \textit{data preparation} steps. Indeed, there is little point in preserving the privacy of clean data sets during model training -- which is currently already possible -- if the raw data has to be leaked first to arrive at those clean data sets! 

In this paper, we contribute to filling this gap in the open literature by proposing \textit{the first MPC based protocol for privacy-preserving feature selection}. Feature selection is the process of selecting a subset of relevant features for model training \cite{CHANDRASHEKAR201416}. Using a well chosen subset of features can lead to more accurate models, as well as efficiency gains during model training. A commonly used technique for feature selection is the so-called \textit{filter method} in which features are ranked according to a score indicative of their predictive ability, and subsequently the highest ranked features are retained. Despite of its known shortcomings, including the fact that it considers each feature in isolation and ignores feature dependencies, the filter method is popular in practical data science because it is computationally very efficient, and  independent of any specific ML model architecture.

The MPC based protocol $\PPFILTERFS$ for private feature selection that we propose in this paper can be used in combination with any MPC protocol to rank features in a privacy-preserving manner. Well-known techniques to score features in terms of their informativeness include mutual information (MI), Gini impurity (GI), and Pearson's correlation coefficient (PCC). 
We propose an efficient feature scoring protocol $\PPGINI$ based on Gini impurity, leaving the development of privacy-preserving protocols for other feature scoring techniques as future work. 
The computation of a GI score for continuous valued features traditionally requires sorting of the feature values to determine candidate split points in the feature value range. As sorting is an expensive operation to perform in a privacy-preserving way, we instead propose a ``mean-split Gini score'' (MS-GINI) that avoids the need for sorting by selecting the mean of the feature values as the split point. As we show in Sec.~\ref{sec:exp}, feature selection with MS-GINI leads to accuracy improvements that are on par with those obtained with GI, PCC, and MI in the data sets used in our experiments. Depending on the application and the data set at hand, one may want to use a different feature scoring technique, in combination with our protocol $\PPFILTERFS$ for private feature selection.

Fig.~\ref{fig:3pc} illustrates the flow of private feature selection and subsequent model training at a high level in an outsourced ``ML as a service setting'' with three computing servers, nicknamed \textit{Alice}, \textit{Bob}, and \textit{Carol} (three-party computation, 3PC). 3PC with honest majority, i.e.~with at most one server being corrupted, is a configuration that is often used in MPC because this setup allows for some of the most efficient MPC schemes. In Step 1 of Fig.~\ref{fig:3pc}, each of $m$ data owners sends secret shares of their data to the three servers (parties). 
While the secret shared data can be trivially revealed by combining shares, no information about the data is revealed by the shares received by any single server, meaning that none of the servers by themselves learn anything about the actual values of the data.
In Step 2A, the three servers execute protocols $\PPGINI$ and $\PPFILTERFS$ to create a reduced version of the data set that contains only the selected features. Throughout this process, none of the parties learns the values of the data or even which features are selected, as all computations are done over secret shares. Next, in Step 2B, the parties train an ML model over the pre-processed data using existing privacy-preserving training protocols, e.g., a privacy-preserving protocol for logistic regression training \cite{BMCMG:CDNRST21}. Finally, in Step 3, the servers can disclose the trained model to the intended model owner by revealing their shares. Steps 1 and 3 are trivial as they follow directly from the choice of the underlying MPC scheme (see Sec.~\ref{sec:smc}). MPC protocols for Step 2B have previously been proposed. The focus of this paper is on Step 2A.
Our approach works in scenarios where the data is horizontally partitioned (each data owner has one or more of the rows or instances), scenarios where the data is vertically partitioned (each data owner has some of the columns or attributes), or any other partition. 

After presenting preliminaries about Gini impurity and MPC in Sec.~\ref{SEC:preliminaries}, and discussing related work in Sec.~\ref{sec:related_work}, we present our main protocol $\PPFILTERFS$ for private feature selection and the supporting protocols $\PPGINIFS$ and $\PPGINI$ in
Sec.~\ref{sec:method}. In Sec.~\ref{sec:exp} we demonstrate the feasibility of our approach for practical data science in terms of accuracy and runtime results through experiments executed on real-world data sets. In our experiments, we consider honest-majority 3PC settings with semi-honest as well as malicious adversaries.
While parties corrupted by semi-honest adversaries follow the protocol instructions correctly but try to obtain additional information, parties corrupted by malicious adversaries can deviate from the protocol instructions. Defending against the latter comes at a higher computational cost which, as we show, can be mitigated by using a recently proposed MPC scheme for 4PC.

\section{Preliminaries}
\label{SEC:preliminaries}

\subsection{Feature Selection based on Gini Impurity}
\label{sec:gini}
 Assume that we have a set $S$ of $m$ training examples, where each training example consists of an input feature vector  $(x_1,\ldots,x_p)$ and a corresponding label $y$. Throughout this paper, we assume that there are $n$ possible class labels. We wish to induce an ML model from this training data that can infer, for a previously unseen input feature vector, a label $y$ as accurately as possible. Not all $p$ features may be equally beneficial to this end. 
 In the filter approach to feature selection, all features are first assigned a score that is indicative of their predictive ability. Subsequently only the best scoring features are retained. A well-known feature scoring criterion is Gini impurity, made popular as part of the classification and regression tree algorithm (CART) \cite{breiman1984}.

If the $j^{th}$ feature $F_j$ is a discrete feature that can assume $\ell$ different values, then it induces a partition $S_1 \cup S_2 \cup \ldots \cup S_\ell$ of $S$ in which $S_i$ is the set of instances that have the $i^{th}$ value for the $j^{th}$ feature. The Gini impurity of $S_i$ is defined as:
\begin{equation}
\label{eq:g_split}
    G(S_i) = \sum_{c=1}^{n} p_c \cdot (1 - p_c)
         = 1 - \sum_{c=1}^{n} p_c^2
\end{equation}
where $p_c$ is the probability of a randomly selected instance from $S_i$ belonging to the $c^{th}$ class.
The Gini score of feature $F_j$ is a weighted average of the Gini impurities of the $S_i$'s:
\begin{equation}
\label{eq:g_attr}
    G(F_j) = \sum_{i=1}^{\ell} \frac{|S_i|}{m} \cdot G(S_i)
\end{equation}

Conceptually, $G(F_j)$ estimates the likelihood of a randomly selected instance to be misclassified based on knowledge of the value of the $j^{th}$ feature. During feature selection, the $k$ features with the lowest Gini scores are retained.

If $F_j$ is a feature with continuous values, then $G(F_j)$ is defined as the weighted average of the Gini impurities of a set $S_{\leq \theta}$ containing all instances for which the $j^{th}$ feature value is smaller than or equal to $\theta$, and a set $S_{> \theta}$ with all instances for which the $j^{th}$ feature value is larger than $\theta$.
In the CART algorithm, an optimal threshold $\theta$ is determined based on sorting of all the instances on their feature values. Since privacy-preserving sorting is a time-consuming operation in MPC \cite{Bogdanov2013ObliviousSO,zigzag}, in Sec.~\ref{sec:msgini} we propose a more straightforward approach for threshold selection which, as we show in Sec.~\ref{sec:exp}, yields desirable improvements in accuracy. 

% section decision_tree_learning (end)

\subsection{Secure Multiparty Computation} % (fold)
\label{sec:smc}
Protocols for MPC enable a set of parties to jointly compute the output of a function over each of the parties' private inputs, without requiring parties to disclose their input to anyone. MPC is concerned with the protocol execution coming under attack by an adversary which may corrupt parties to learn private information or cause the result of the computation to be incorrect. MPC protocols are designed to prevent such attacks being successful, and use proven cryptographic techniques to guarantee privacy.

\textbf{Adversarial Model:} An adversary $\mathcal{A}$ can corrupt any number of parties. In a \textit{dishonest-majority} setting, half or more of the parties may be corrupt, while in an \textit{honest-majority} setting, more than half of the parties are honest (not corrupted). Furthermore, $\mathcal{A}$ can be a \textit{semi-honest} or a \textit{malicious} adversary. While a party corrupted by a semi-honest or ``passive'' adversary follows the protocol instructions correctly but tries to obtain additional information, parties corrupted by malicious or ``active'' adversaries can deviate from the protocol instructions. 
The protocols in Sec.~\ref{sec:method} are sufficiently generic to be used in dishonest-majority as well as honest-majority settings, with
passive or active adversaries. This is achieved by changing
the underlying MPC scheme to align with the desired
security setting. Some of the most efficient MPC schemes have been developed for 3 parties, out of which at most one is corrupted. We evaluate the runtime of our protocols in this honest-majority 3PC  setting, which is growing in popularity in the PPML literature, e.g.~\cite{dalskov2019secure,kumar2020cryptflow,riazi2018chameleon,wagh2019securenn}, and we demonstrate how even better runtimes can be obtained with a recently proposed MPC scheme for 4PC with one corruption \cite{cryptoeprint:2020:1330}.

In the MPC schemes used in this paper, all computations by the parties (servers) are done over integers in a ring $\mathbb{Z}_{q}$. Raw data in ML applications is often real-valued. As is common in the MPC literature, we convert real numbers to integers using a fixed-point representation \cite{fixed_point_conversion}.
After this conversion, the data owners secret share their values with the parties using a secret sharing scheme and proceed by performing operations over the secret shares.

%We consider the three-party secure computation (3PC) setting; the adversary can corrupt at most one server. Because this setting allows for efficient protocols, the honest-majority 3PC setting is growing in popularity in the PPML literature (e.g.~\cite{bittner2020private,riazi2018chameleon,wagh2019securenn}).

For the \textbf{passive 3PC} setting, we follow a \textit{replicated secret sharing} scheme from Araki et al.~(\cite{10.1145/2976749.2978331}).
% After this conversion, in the \textit{semi-honest} adversary setting, the data owners secret share their values with the parties using a replicated secret sharing scheme \cite{10.1145/2976749.2978331}.
To share a secret value $x \in \mathbb{Z}_{q}$ among parties $P_1, P_2$ and $P_3$, the shares $x_1, x_2, x_3$ are chosen uniformly at random in $\mathbb{Z}_{q}$ with the constraint that $x_1 + x_2 +x_3 = x \mod q$. $P_1$ receives $x_1$ and $x_2$, $P_2$ receives $x_2$ and $x_3$, and $P_3$ receives $x_3$ and $x_1$. Note that it is necessary to combine the shares available to two parties in order to recover $x$, and no information about the secret shared value $x$ is revealed to any single party. For short, we denote this secret sharing by $[\![x]\!]_q$. Let $[\![x]\!]_q$, $[\![y]\!]_q$ be secret shared values and $c$ be a constant, the following computations can be done locally by parties without communication:
\begin{itemize}[leftmargin=*,noitemsep,topsep=0pt]
  \item Addition ($z=x+y$): Each party $P_i$ gets shares of $z$ by computing
  $z_i=x_i+y_i$ and $z_{(i + 1 {\rm \ mod\ } 3) }$ $=$ $x_{(i + 1 {\rm \ mod\ } 3) }$ $+$ $y_{(i + 1 {\rm \ mod\ } 3) }$. This is denoted by 
  $[\![z]\!]_q \leftarrow[\![x]\!]_q+[\![y]\!]_q$.
  \item Subtraction $[\![z]\!]_q\leftarrow[\![x]\!]_q-[\![y]\!]_q$ is performed analogously.

  \item Multiplication by a constant ($z=c \cdot x$): Each party multiplies its local shares of $x$ by $c$ to obtain shares of $z$. This is denoted by 
  $[\![z]\!]_q\leftarrow c \cdot [\![x]\!]_q$
  \item Addition of a constant ($z=x+c$): $P_1$ and $P_3$ add $c$ to their share $x_1$ of $x$ to obtain $z_1$, while the parties set $z_2=x_2$ and $z_3=x_3$. This will be denoted by $[\![z]\!]_q\leftarrow[\![x]\!]_q + c$.
\end{itemize}

The main advantage of replicated secret sharing 
compared to other secret sharing schemes
is that replicated shares enables a very efficient procedure for multiplying secret shared values. To compute $x \cdot y=(x_1 + x_2 +x_3)(y_1 + y_2 +y_3)$, the parties  
locally perform the following computations: $P_1$ computes $z_1$ $=$ $x_1 \cdot y_1$ $+$ $x_1 \cdot y_2$ $+$ $x_2 \cdot y_1$, $P_2$ computes $z_2$ $=$ $x_2 \cdot y_2$ $+$ $x_2 \cdot y_3$ $+$ $x_3 \cdot y_2$ and $P_3$ computes $z_3$ $=$ $x_3 \cdot y_3$ $+$ $x_3 \cdot y_1$ $+$ $x_1 \cdot y_3$. By doing so, without any interaction, each $P_i$ obtains $z_i$ such that $z_1+z_2+z_3 = x \cdot y \mod q$.
After that, the parties are required to convert from this additive secret sharing representation back to the original replicated secret sharing representation (which requires that the parties add a secret sharing of zero and that each party sends one share to one other party for a total communication of three shares). See \cite{10.1145/2976749.2978331} for more details. 

In the \textbf{active 3PC} setting,
we use the MPC scheme \textit{SYReplicated2k} recently proposed by Dalskov et al.~(\cite{cryptoeprint:2020:1330}).
%with the option with preprocessing for generation of the multiplication triples that is available in MP-SPDZ \cite{cryptoeprint:2020:521}.
%an MPC scheme that was recently proposed by \cite{Abspoel2019AnEP}.
%Eerikson et al.~(\cite{eerikson2020use}). 
%In the \textit{malicious} adversary setting, 
In this MPC scheme, the parties are prevented from deviating from the protocol and from gaining knowledge from other parties through the use of information-theoretic message authentication codes (MACs). 
In addition to computations over secret shares of the data, the parties also perform computations required for MACs. 
%The operations are slightly more involved than in the semi-honest security setting, and the total communication cost is higher.  
%See \cite{eerikson2020use} for details.
See \cite{cryptoeprint:2020:1330} for details.
Finally, we use the MPC scheme recently proposed by Dalskov et al.~(\cite{cryptoeprint:2020:1330})
for the \textbf{active 4PC} setting, where the computations are outsourced to four servers out of which at most one has been corrupted by a malicious adversary.

\textbf{Building Blocks:} 
Building on the cryptographic primitives listed above for addition and multiplication of secret shared values, MPC protocols for other operations have been developed in the literature. In this paper, we use: 
\begin{itemize}[leftmargin=*,noitemsep,topsep=0pt]
    \item Secure matrix multiplication \pmmul: at the start of this protocol, the parties have secret sharings $[\![A]\!]$ and $[\![B]\!]$ of matrices $A$ and $B$; at the end, the parties have a secret sharing $[\![C]\!]$ of the product of the matrices, $C=A \times B$. \pmmul can be constructed as a direct extension of the secure multiplication protocol for two integers, which we will denote as \pmul in the remainder of the paper. Similarly, we use $\pDP$ to denote the protocol for the secure dot product of two vectors. In a replicated sharing scheme, dot products can be computed more efficiently than the direct extension from $\pmul$, and matrix multiplication can use this optimized version of dot products; we refer to Keller (\cite{cryptoeprint:2020:521}) for details.
    
    \item Secure comparison protocol \plt \cite{catrina2010improved}: at the start of this protocol, the parties have secret sharings $[\![x]\!]$ and $[\![y]\!]$ of two integers $x$ and $y$; at the end, they have a secret sharing of $1$ if $x < y$, and a secret sharing of $0$ otherwise.
    
    \item Secure argmin protocol \pargmin : this protocol accepts secret sharings of a vector of integers and returns a secret sharing of the index at which the vector has the minimum value. \pargmin is straightforwardly constructed using the above mentioned secure comparison protocol.
    
    \item Secure equality test protocol $\peq$ \cite{fixed_point_conversion}: at the start of this protocol, the parties have secret sharings $[\![x]\!]$ and $[\![y]\!]$ of two integers $x$ and $y$; at the end, they have a secret sharing of $1$ if $x = y$, and a secret sharing of $0$ otherwise.
    
    \item Secure division protocol $\pDIV$ \cite{fixed_point_conversion}: at the start of this protocol, the parties have secret sharings $[\![x]\!]_q$ and $[\![y]\!]_q$ of two integers $x$ and $y$; at the end, they have a secret sharing $[\![z]\!]_q$ of %the division of $x$ and $y$, 
    $z = x / y$.

\end{itemize}

\section{Related Work}
\label{sec:related_work}
\textbf{Private Feature Selection:}
Given that feature selection is an important step in the data preparation pipeline, it has received remarkably little attention in the PPML literature to date. Feature selection techniques have been proposed that favor features that do not contain sensitive information \cite{jafer2015framework}. Work like that is orthogonal to ours, as it assumes the existence of a data curator with full access to all the data. 
Regarding approaches to private feature selection among multiple data owners, early attempts
\cite{banerjee2011privacy,sheikhalishahi2017privacy} in the semi-honest setting use a ``distributed secure sum protocol'' reminiscent of the way in which sums are computed in MPC based on secret sharing (see Sec.~\ref{sec:smc}). The limitations of this work in terms of security include the fact that the parties find out which features are selected, and 
statistical information about the data is leaked to all parties during the computation of the feature scores, as only summations, and not other operations, are done in a secure manner.
~\cite{Rao2019SecureTF} proposed a more principled 2PC protocol with Paillier homomorphic encryption for private feature selection with $\chi^2$ as filter criteria in the semi-honest setting, without an experimental evaluation of the proposed approach. To the best of our knowledge, private feature selection with malicious adversaries has not yet been proposed or evaluated.  
The recent approach by ~\cite{ye2019distributed} is not based on cryptography, does not provide any formal privacy guarantees, and leaks information through disclosure of intermediate representations.

\textbf{Secure Gini Score Computation:}
Besides as a technique to score features for feature selection, as we do in this paper, Gini impurity is traditionally used in ML in the CART algorithm for training decision trees \cite{breiman1984}, and it has been adopted in MPC protocols for privacy-preserving training of decision tree models \cite{de2014practical,Choudhary:2020,escudero2020}.
Gini score computation for continuous valued features, as we do in this paper, is especially challenging from an MPC point of view, as it requires sorting of feature values to determine candidate split points in the feature range. Abspoel et al.~(\cite{escudero2020}) put ample effort in performing this sorting process as efficiently as possible in a secure manner. We take a drastically different approach by assuming that the mean of the feature values serves as a good approximation for an optimal split threshold. This has the double advantage that (1) there is no need for oblivious sorting of feature values, and (2) for each feature only one Gini score for one threshold $\theta$ has to be computed as opposed to computing the Gini score for multiple candidate thresholds and then selecting the best one through secure comparisons. This leads to significant efficiency gains, while preserving good accuracy, as we demonstrate in Sec.~\ref{sec:exp}.

\begin{protocol}
    \caption{Protocol \PPFILTERFS for Secure Filter based Feature Selection}
    \label{prtc:filter_fs}
    
    \textbf{Input:} A secret shared $m \times p$ data matrix $[\![D]\!]_q$, a secret shared $p$-length score vector $[\![G]\!]_q$, the number $k < p$ of features to be selected, and a constant $t$ that is bigger than the highest possible score in $[\![G]\!]_q$
    
    \textbf{Output:} a secret shared $m \times k$ matrix $[\![D']\!]_q$ 
    
\begin{algorithmic}[1]
    \FOR{$i=1$ {\bfseries to} $k$} 
        \STATE $[\![I[i]]\!]_q \gets$ \pargmin $([\![G]\!]_q)$
    
        \FOR{$j \gets 1$ {\bfseries to} $p$}
            \STATE $[\![flag_k]\!]_q \gets$ \peq $([\![I[i]]\!]_q, j)$ \\
            
            \STATE $[\![T[j][i]]\!]_q \gets [\![flag_k]\!]_q$ \\
            
            \STATE $[\![G[j]]\!]_q \gets [\![G[j]]\!]_q + $ \pmul $([\![flag_k]\!]_q, t - [\![G[j]]\!]_q)$
        \ENDFOR
    \ENDFOR
    
    \STATE $[\![D']\!]_q\gets$ \pmmul $([\![D]\!]_q, [\![T]\!]_q)$ \\
    
    \STATE \textbf{return} $[\![D']\!]_q$
\end{algorithmic}
\end{protocol}

\section{Methodology}
\label{sec:method}
We present a protocol for oblivious feature selection based on precomputed scores for the features, followed by a protocol for computing the feature scores themselves in a private  manner. 
In Sec.~\ref{sec:exp} we evaluate the protocols in 3PC and 4PC honest-majority settings.
%with parties \textit{Alice}, \textit{Bob}, and \textit{Carol}.

\subsection{Secure Filter based Feature Selection}
\label{sec:filterfs}

At the start of the Protocol $\PPFILTERFS$ for secure feature selection, the parties have secret shares of a data matrix $D$ of size $m \times p$, in which the rows correspond to instances and the columns to features. The parties also have secret shares of a vector $G$ of length $p$ containing a score for each of the features. At the end of the protocol, the parties have a reduced matrix $D'$ of size $m \times k$ in which only the columns from $D$ corresponding to the lowest scores in $G$ are retained (note that this protocol can be trivially modified to select the $k$ features with the highest scores). 
The main ideas behind the protocol (which is described in Protocol \ref{prtc:filter_fs}) are to:
\begin{enumerate}[leftmargin=*,noitemsep,topsep=0pt]
\item Determine the indices of the features that need to be selected (these are stored in a secret-shared way in $I$).
\item Create a matrix $T$ in which the columns are one-hot-encoded representations of these indices.
\item Multiply $D$ with this feature selection matrix $T$. 
\end{enumerate}

\noindent
Before walking through the pseudocode of Protocol \ref{prtc:filter_fs}, we present a plaintext example to illustrate the notation.

%\begin{example}\label{EX:FEATSEL}
\textbf{Example 1.} Consider the data matrix $D$ at the left of Equation (\ref{EQ:FEATSEL}), containing values for $m=5$ instances (rows) and $p=4$ features (columns). Assume that the feature score vector is $G=[65, 26, 83, 14]$ and that we want to select the $k=2$ features with the lowest scores in $G$.
\begin{equation}\label{EQ:FEATSEL}
\scriptsize{
    \underbrace{
    \left(
    \begin{array}{cccc}
    1 & 2 & 3 & 4\\
    5 & 6 & 7 & 8\\
    9 & 10 & 11 & 12\\
    13 & 14 & 15 & 16\\
    17 & 18 & 19 & 20
    \end{array}
    \right)}_{D} \cdot
    \underbrace{
     \left(
    \begin{array}{cc}
    0 & 0\\
    0 & 1\\
    0 & 0\\
    1 & 0
    \end{array}
    \right)}_{T} = 
    \underbrace{ 
    \left(
    \begin{array}{cc}
    4 & 2\\
    8 & 6\\
    12 & 10\\
    16 & 14\\
    20 & 18 
    \end{array}
    \right)}_{D'}
    }
\end{equation}
The lowest scores in $G$ are 14 and 26, hence the 4th and the 2nd column of $D$ should be selected. The columns of $T$ in Equation (\ref{EQ:FEATSEL}) are a one-hot-encoding of 4 and 2 respectively, and multiplying $D$ with $T$ will yield the desired reduced data matrix $D'$. This multiplication takes place on Line 9 in Protocol \ref{prtc:filter_fs}. The bulk of Protocol \ref{prtc:filter_fs} is about how to construct $T$ based on $G$. As explained below, this process involves an auxiliary vector, which, at the end of the protocol, contains the following values for our example:
%$G_{least} = [14,26]$ and 
$I=[4,2]$.
%\end{example}

In the protocol, vector $[\![I]\!]_q$ of length $k$ stores the indices of the $k$ selected features out of the $p$ features of $[\![D]\!]_q$ and matrix $[\![T]\!]_q$ is a $p \times k$ transformation matrix that eventually holds one-hot-encodings of the indices in $I$.
Through executing Lines 1-8 of Protocol \ref{prtc:filter_fs}, the parties construct a feature selection matrix $T$ based on the values in $G$.
%is about how we select $k$ features based on computed scores $[\![G]\!]_q$ %and is also another example of how to apply the trick in Protocol \ref{prtc:ms_gini} to update variables instead of if-else statements. 
%The first inner for-loop, on Line 3-7, identifies the $i^{th}$ smallest value in $[\![G]\!]_q$ (which ends up stored in $[\![G_{least}[i]]\!]_q$ of Line 6) and the index of this value in $[\![G]\!]_q$ (which ends up stored in $[\![I[i]]\!]_q$ of Line 5). The second inner for-loop, on Line 8-12, updates the transformation matrix $[\![T]\!]_q$ (Line 10) and marks the identified feature as ``selected'' (Line 11).
In Line 2 the index of the $i^{th}$ smallest value in $[\![G]\!]_q$ is identified. To this end, the parties run a secure argmin protocol $\pargmin$. 
%and store it in $[\![I[i]]\!]_q$. 
The inner for-loop serves two purposes, namely constructing the $i^{th}$ column of matrix $T$, and overwriting the score in $G$ of the feature that was selected in Line 2 by the upper bound, so that it will not be selected anymore in further iterations of the outer for-loop (such an upper bound $t$ is passed as input to Protocol \ref{prtc:filter_fs} and is usually very easy to determine in practice, as most common feature scoring techniques range between $0$ and $1$):
\begin{itemize}[leftmargin=*,noitemsep,topsep=0pt]
\item To construct the $i^{th}$ column of $T$, the parties loop through row $j=1 \ldots p$, and on Line 5, update $T[j][i]$ with either a 0 or a 1, depending on the outcome of the secure equality test on Line 4. The outcome of this test will be 1 exactly once, namely when $j$ equals $I[i]$, hence Line 5 results in a one-hot-encoding of $I[i]$ stored in the $i$th column of $T$. 
\item The flag $flag_{k}$ computed on Line 4 is used again on Line 6 to overwrite $G[I[i]]$ with $t$ in an oblivious manner, where $t$ is a value that is larger than the highest possible score that occurs in $[\![G]\!]_q$. This theoretical upper bound $t$ ensures that feature $I[i]$ will not be selected again in later iterations of the outer for-loop. 
\end{itemize}

As is common in MPC protocols, we use multiplication instead of control flow logic for conditional assignments. To this end, a conditional based branch operation as ``$\textbf{if  } c \textbf{  then  } a \leftarrow b$'' is rephrased as 
%\begin{equation}
$a  \gets  a  +  c  \cdot  (b- a)$.
%\end{equation}
In this way, the number and the kind of operations executed by the parties does not depend on the actual values of the inputs, so it does not leak information that could be exploited by side-channel attacks. Such a conditional assignment occurs in Line 6 of Protocol \ref{prtc:filter_fs}, where the value of the condition $c$ itself is computed on Line 4.
%
%Within the first inner for-loop, a condition is checked, namely whether the current score $G[j]$ is smaller than the smallest one seen so far in this iteration of the outer for-loop by secure comparison protocol $\plt$: 
%\magenta{I think that the standard comparison protocol checks $\geq$ and not $<$ so you need to be careful how you phrase it.}
%\begin{equation}
%[\![G[j]]\!]_q < [\![G_{least}[i]]\!]_q
%\plt([\![G[j]]\!]_q, [\![G_{least}[i]]\!]_q)
%\end{equation}
%and whether the feature under consideration has not been selected before, i.e.
%\begin{equation}
%\label{eq:selected_or_not}
%1 - [\![S[j]]\!]_q
%\end{equation}
%Both checks are combined by secure multiplication protocol $\pmul$ in Line 5.
%and $[\![S[j]]\!]_q == [\![0]\!]_q$ shown in \ref{eq:flag_lf}. 
%In each iteration of first inner for-loop, we want to update $[\![I[i]]\!]_q$ and $[\![G_{least}[i]]\!]_q$ (select feature $j$) when we find that the current value of $[\![G[j]]\!]_q$ is less than the value of $[\![G_{least}[i]]\!]_q$ in the record 
%($[\![flag_{l}]\!]_q = [\![1]\!]_q$) and the feature $j$ is not selected in previous steps ($[\![S[j]]\!]_q = [\![0]\!]_q$)
%. That means we update $[\![I[i]]\!]_q$ and $[\![G_{least}[i]]\!]_q$ by $j$ and $[\![G[j]]\!]_q$ respectively when $[\![flag_{l}]\!]_q = [\![1]\!]_q$. Otherwise, $[\![I[i]]\!]_q$ and $[\![G_{least}[i]]\!]_q$ are not changed.
%
%
In the final step, on Line 9, the parties multiply matrix $D$ with matrix $T$ in a secure manner to obtain a matrix $D'$ that contains only the feature columns corresponding to the $k$ best features. Throughout this process, the parties are unaware of which features were actually selected. The secret shared matrix $D'$ can subsequently be used as input for a privacy-preserving ML model training protocol, e.g.~\cite{BMCMG:CDNRST21}.

%Because we do not want to leak information of selected indices $[\![I]\!]_q$, feature selection is actually done by matrix multiplication (MATMUL) of $[\![D]\!]_q$ with $[\![T]\!]_q$. 
%and $[\![D']\!]_q$ is a $m \times k$ matrix consisted of all k selected features. 
%Each row of $[\![T]\!]_q$ represents each feature of $[\![D]\!]_q$ and each column of $[\![T]\!]_q$ represents each feature of $[\![D']\!]_q$, so $[\![T[j][i]]\!]_q = [\![1]\!]_q$ means $j^{th}$ feature of $[\![D]\!]_q$ is selected as $i^{th}$ feature of $[\![D']\!]_q$.

%Moreover, for feature selection part in in-the-clear version, it is easy for us to pick the columns corresponding to the selected features and form $D'$ directly by accessing the selected indices in $D$. To select features and form $[\![D']\!]_q$ in privacy-preserving way, we use the transformation matrix $[\![T]\!]_q$ that is constructed in an oblivious manner on Line 15. Multiplying the original data matrix $[\![D]\!]_q$ with $[\![T]\!]_q$ allows to select the desired feature columns while hiding information about the selected indices by not accessing specific locations of $[\![D]\!]_q$ only. We denote the result of MATMUL of $[\![D]\!]_q$ and $[\![T]\!]_q$ as $[\![D']\!]_q$ and secure MATMUL is computed by protocol $\pmmul$.

\subsection{Secure Feature Score Computation}
\label{sec:msgini}
Protocol $\PPFILTERFS$ assumes the availability of a feature score vector $G$ and an upper bound $t$ for the values in $G$. Below we explain how this can be obtained from the data in a secure manner. To this end, we present a protocol $\PPGINI$ for computation of the score of a feature based on Gini impurity. This protocol is applicable to data sets with continuous features. It is computationally cheaper than previously proposed protocols for Gini impurity that rely on sorting of feature values. Furthermore, as shown in previous work \cite{cogLoad} and in Sec.~\ref{sec:exp}, the ``Mean-Split'' GINI score can yield similar accuracy improvements.  

Recall that we have a set $S$ of $m$ training examples, where each training example consists of an input feature vector  $(x_1,\ldots,x_p)$ and a corresponding label $y$. We propose to split the set of values of the $j^{th}$ feature $F_j$ based on its mean value as a threshold $\theta$. We denote by $S_{\leq \theta}$ the set of instances that have $x_j \leq \theta$, and by $S_{>\theta}$ the set of instances that have $x_j > \theta$. Furthermore, for $c = 1, \ldots, n$, we denote by $L_c$ the set of examples from $S$ that have class label $y = c$.
Based on the binary split, we define the 
MS-GINI (``Mean-Split'' GINI) score for feature $F_j$ as:
\begin{equation}
\label{GINI1}
G(F_j) 
 =  \frac{1}{m} \cdot (|S_{\leq \theta}| \cdot G(S_{\leq \theta}) + |S_{>\theta}| \cdot G(S_{>\theta}))
\end{equation}
with the Gini impurities of $S_{\leq \theta}$ and $S_{> \theta}$ defined as:
\begin{equation}
\label{GINI2}
G(S_{\leq \theta}) = 1 - \sum\limits_{c=1}^n (p_c^{\leq \theta})^2\mbox{;\ \ }
\displaystyle G(S_{>\theta}) = 1 - \sum\limits_{c=1}^n (p_c^{> \theta})^2
\end{equation}
and the probabilities defined as:
\begin{equation}
\label{GINI3}
p_c^{\leq \theta} = \frac{|S_{\leq \theta} \cap L_c|}{|S_{\leq \theta}|}\mbox{;\ \ \ }
p_c^{> \theta} = \frac{|S_{> \theta} \cap L_c|}{|S_{> \theta}|}
\end{equation}
Formulas (\ref{GINI1}), (\ref{GINI2}) and (\ref{GINI3})
are consistent with the definition of Gini score given in Sec.~\ref{SEC:preliminaries}, and presented here in more detail to enhance the readability of our secure protocol $\PPGINI$ for the computation of the Gini score $G(F)$ of feature $F$ (described in Protocol \ref{prtc:ms_gini}).

\begin{protocol}
    \caption{Protocol $\PPGINI$ for Secure MS-GINI Score of a Feature}
    \label{prtc:ms_gini}

    \textbf{Input:} A secret shared feature column $[\![F]\!]_q$ = ($[\![f_1]\!]_q$,$[\![f_2]\!]_q$,...,$[\![f_m]\!]_q$), a secret shared $m \times (n - 1)$ label-class matrix $[\![L]\!]_q$, where $m$ is the number of instances and $n$ is the number of classes.
    
    \textbf{Output:} MS-GINI score $[\![G(F)]\!]_q$ of the feature $F$
    
    \begin{algorithmic}[1]
    
    \STATE $[\![\theta]\!]_q \gets ([\![f_1]\!]_q + [\![f_2]\!]_q + ... + [\![f_m]\!]_q) \cdot \frac{1}{m}$ 
    
    \STATE Initialize $[\![a]\!]_q$, $[\![b]\!]_q$, $[\![A]\!]_q$ and $[\![B]\!]_q$ with zeros.
    
    \FOR{$i\gets 1$ {\bfseries to} $m$}
        
        \STATE $[\![flag_s]\!]_q \gets \plt([\![\theta]\!]_q,[\![f_i]\!]_q)$
        
        \STATE $[\![b]\!]_q \gets [\![b]\!]_q + [\![flag_s]\!]_q$ 
        
        \FOR{$j\gets 1$ {\bfseries to} $n - 1$}
            \STATE $[\![flag_m]\!]_q \gets \pmul([\![flag_s]\!]_q,[\![L[i][j]]\!]_q)$
            
            \STATE $[\![B[j]]\!]_q \gets [\![B[j]]\!]_q + [\![flag_m]\!]_q$
            
            \STATE $[\![A[j]]\!]_q \gets [\![A[j]]\!]_q + [\![L[i][j]]\!]_q - [\![flag_m]\!]_q$
        \ENDFOR
    \ENDFOR
    \STATE $[\![a]\!]_q \gets m - [\![b]\!]_q$
    
    \STATE $[\![A[n]]\!]_q \gets [\![a]\!]_q - ([\![A[1]]\!]_q + ... + [\![A[n-1]]\!]_q)$
    
    \STATE $[\![B[n]]\!]_q \gets [\![b]\!]_q - ([\![B[1]]\!]_q + ... + [\![B[n-1]]\!]_q)$
    
    \STATE $[\![G(\SLT)]\!]_q \gets [\![a]\!]_q - $ \pmul $($ \pDP $([\![A]\!]_q,[\![A]\!]_q),$ \pDIV $(1,[\![a]\!]_q))$
    
    \STATE $[\![G(\SGT)]\!]_q \gets [\![b]\!]_q - $ \pmul $($ \pDP $([\![B]\!]_q,[\![B]\!]_q),$ \pDIV $(1,[\![b]\!]_q))$
    
    \STATE $[\![G(F)]\!]_q \gets [\![G(\SLT)]\!]_q + [\![G(\SGT)]\!]_q$
    
    \STATE \textbf{return} $[\![G(F)]\!]_q$
    \end{algorithmic}

\end{protocol}

At the start of Protocol $\PPGINI$, the parties have secret shares of a feature column $F$ (think of this as a column from data matrix $D$ in Example 1), as well as secret shares of an one-hot-encoded version of the label vector. The latter is represented as a label-class matrix $[\![L]\!]_q$, in which $[\![L[i][j]]\!]_q = [\![1]\!]_q$ means that the label of the $i^{th}$ instance is equal to the $j^{th}$ class. Otherwise, $[\![L[i][j]]\!]_q = [\![0]\!]_q$. We note that, while there are $n$ classes, it is sufficient for $L$ to contain only $n-1$ columns: as there is exactly one value 1 per row, the value of the $n^{th}$ column is implicit from the values of the other columns. We indirectly take advantage of this fact by terminating the loop on Line 6-10 at $n-1$, and performing calculations for the $n^{th}$ class separately and in a cheaper manner on Line 13-14, as we explain in more detail below. 

On Line 1, the parties compute $[\![\theta]\!]_q$ as a threshold to split the input feature $[\![F]\!]_q$, as the mean of the feature values in the column. To this end, each party first sums up the secret shares of the feature values, and then multiplies the sum with a known constant $\frac{1}{m}$ locally.
Line 2 is to initialize all counters related to $\SLT$ and $\SGT$ to zero. After Line 14, these counters will contain the following values:
$$
\begin{array}{rcl}
a & = & |S_{\leq \theta}|\\
b & = & |S_{> \theta}|\\
A[j] & = & |S_{\leq \theta} \cap L_j|\mbox{,\ for\ }j=1 \ldots n\\
B[j] & = & |S_{> \theta} \cap L_j|\mbox{,\ for\ }j=1 \ldots n\\
\end{array}
$$
%$[\![a]\!]_q$ and $[\![b]\!]_q$ are counters to record the number of instances in $S_{\leq \theta}$ and $S_{> \theta}$, i.e.~at the end of the protocol $a = |S_{\leq \theta}|$ and $b=|S_{> \theta}|$. $[\![A]\!]_q$ and $[\![B]\!]_q$ are vectors of counters with length n. $[\![A[i]]\!]_q$ is the counter to record number of instances from $S_{\leq \theta}$ with corresponding label being equal to $i^{th}$ class and $[\![B[i]]\!]_q$ is the counter to record number of instances from $S_{> \theta}$ with corresponding label being equal to $i^{th}$ class. 
These counters are needed for the probabilities in Equation (\ref{GINI3}).
For each instance, in Line 4 of Protocol \ref{prtc:ms_gini}, the parties perform a secure comparison to determine whether the instance belongs to $S_{> \theta}$. The outcome of that test is added to $b$ on Line 5. Since the total number of instances is $m$, $a$ can be straightforwardly computed as $m-b$ after the outer for-loop, i.e.~on Line 12.
Lines 7-8 check whether the instance belongs to $S_{> \theta} \cap L_j$, in which case $B[j]$ is incremented by 1. The equivalent operation of Line 7-8 for $A[j]$ would be
$[\![A[j]]\!]_q \gets [\![A[j]]\!]_q + \pmul((1 - [\![flag_s]\!]_q),[\![L[i][j]]\!]_q)$.
We have simplified this instruction on Line 9, taking advantage of the fact that $\pmul([\![flag_s]\!]_q,[\![L[i][j]]\!]_q)$ has been precomputed as $[\![flag_m]\!]_q$ on Line 7.

On Line 13-14 the parties compute $[\![A[n]]\!]_q$ and $[\![B[n]]\!]_q$, leveraging the fact that sum of all values in $[\![A]\!]_q$ is $[\![a]\!]_q$, and the sum of all values in $[\![B]\!]_q$ is $[\![b]\!]_q$.
All operations on Line 13-14 can be performed locally by the parties, on their own shares. Moving the computation of $[\![A[n]]\!]_q$ and $[\![B[n]]\!]_q$ out of the for-loop, reduces the number of secure multiplications needed from $m \times n$ to $m \times (n-1)$. In the case of a binary classification problem, i.e.~$n=2$, this means that the number of secure multiplications required is cut down by half.

%In this way, we can save runtime by reducing one-iteration communication of $\pmul$ among parties and let parties to compute $[\![A[n]]\!]_q$ and $[\![B[n]]\!]_q$ locally. If $n=2$, it is a binary classification problem and we do not even have inner for-loop.

Using the notations for the counters from the pseudocode of Protocol \ref{prtc:ms_gini}, Equation (\ref{GINI1}) comes down to:\\

\noindent
{\scriptsize
$G(F)  =   \displaystyle \frac{1}{m} \cdot \left[
a \cdot \left( 1 - \sum\limits_{j=1}^n \left(\frac{A[j]}{a}\right)^2 \right) \right. 
 \displaystyle +
\left. b \cdot \left( 1 - \sum\limits_{j=1}^n \left(\frac{B[j]}{b}\right)^2 \right) \right]$\\
$\phantom{G(F)} =   
\displaystyle \frac{1}{m} \cdot \left[
\left(a - \frac{1}{a} \cdot A \bullet A \right)
+
\left(b - \frac{1}{b} \cdot B \bullet B \right) \right]
$
}\\

\noindent
in which $A \bullet A$ and $B \bullet B$ are the dot products of $A$ and $B$ with themselves, respectively. These computations are performed by the parties on Lines 15-17 using, among other things, the protocol $\pDP$ for secure dot product of vectors, and the protocol $\pDIV$ for secure division. We note that the final multiplication with the factor $1/m$ is omitted altogether from Protocol \ref{prtc:ms_gini} as this will have no effect on the relative ordering of the scores of the individual features.

If data are vertically partitioned and all data owners have the label vector, they can compute MS-GINI scores offline without $\PPGINI$, and the computing servers would only have to do feature selection based on  pre-computed MS-GINI scores with Protocol $\PPFILTERFS$. In reality, often, it is not reasonable to allow each data owner to have all labels, so we do not assume this scenario in our protocols.

\subsection{Secure Feature Selection with MS-GINI}
\label{sec:ginifs}

Protocol $\PPGINIFS$ (described in Protocol \ref{prtc:msgini_fs}) performs secure filter-based feature selection with MS-GINI, used for the experiments in this work. It %relies on both $\PPGINI$ and $\PPFILTERFS$, and as such 
combines the building blocks presented earlier in the section. 
By executing the loop on Line 1-3, the parties compute the MS-GINI score of the $i^{th}$ feature from the original data matrix $[\![D]\!]_q$ using Protocol $\PPGINI$, and store it into $[\![G[i]]\!]_q$. On
Line 4, the parties perform filter-based feature selection using Protocol $\PPFILTERFS$ to obtain a $m \times k$ matrix $[\![D']\!]_q$ with $k$ selected features from $[\![D]\!]_q$.
As the standard GINI score is upper bounded by 1, and $\PPGINI$ ignores the multiplication by $1/m$ for efficiency reasons, it is safe to use $m$ as the upper bound that is passed to Protocol $\PPFILTERFS$ on Line 4.

\begin{protocol}
    \caption{Protocol $\PPGINIFS$ for Secure Filter-based Feature Selection with MS-GINI}
    \label{prtc:msgini_fs}
    
    \textbf{Input:} A secret shared $m \times p$ data matrix $[\![D]\!]_q$ = ($[\![F_1]\!]_q$,$[\![F_2]\!]_q$,...,$[\![F_p]\!]_q$), a secret shared $m \times (n - 1)$ label-class matrix $[\![L]\!]_q$, where $m$ is
    the number of instances, $p$ the number of features, $n$ the number of classes, and $k$ the number of features to be selected.
    
    \textbf{Output:} a secret shared $m \times k$ matrix $[\![D']\!]_q$
    
    \begin{algorithmic}[1]
    \FOR{$i\gets 1$ {\bfseries to} $p$}
        \STATE $[\![G[i]]\!]_q \gets \PPGINI([\![F_i]\!]_q,[\![L]\!]_q,m,n)$
    \ENDFOR
    
    \STATE $[\![D']\!]_q \gets \PPFILTERFS([\![D]\!]_q,[\![G]\!]_q,k,m)$
    
    \STATE \textbf{return} $[\![D']\!]_q$
    \end{algorithmic}
\end{protocol}

\begin{table*}
    \caption{Feature selection accuracy and runtime results}
    \label{table:acc_tab}
    \centering
    \scriptsize{
    \begin{tabular}{|c|c|c|c|c|c|c|c|c|c|r|r|r|}
    \cline{1-13}
    \multicolumn{1}{|c|}{} & \multicolumn{4}{|c|}{data set details} & \multicolumn{5}{|c|}{logistic regression accuracy results} & \multicolumn{3}{|c|}{runtime} \\
    %\multicolumn{2}{|c|}{semi-honest} & \multicolumn{2}{|c|}{malicious}\\ 
    \hline
    Data set & $m$ & $p$ & $k$ & \#folds & RAW & MS-GINI & GI & PCC & MI & \multicolumn{1}{|c|}{passive 3PC} & \multicolumn{1}{|c|}{active 3PC} &  \multicolumn{1}{|c|}{active 4PC} \\
    %$\mathbb{Z}_p$ & $\mathbb{Z}_{2^l}$ & $\mathbb{Z}_p$ & $\mathbb{Z}_{2^l}$\\
    \hline
    CogLoad & 632 & 120 & 12 & 6 & 50.90\% & 52.50\% & 52.70\% & 48.57\% & 51.59\% & 50 sec & 163 sec & 79 sec \\
    \hline
    LSVT & 126 & 310 & 103 & 10 & 80.09\% & 86.15\% & 82.74\% & 78.89\% & 85.38\% & 60 sec & 254 sec & 89 sec \\
    \hline
    SPEED & 8,378 & 122 & 67 & 10 & 95.24\% & 97.26\% & 95.56\% & 95.89\% & 95.83\% & 949 sec & 3,634 sec & 1,435 sec \\
    \hline
    \end{tabular}
    }
\end{table*}

\begin{table}
    \caption{Runtime details for active 3PC}
    \label{table:perf_tab}
    \centering
    \scriptsize{
    \begin{tabular}{|c|c|c|c|r|r|r|}
    \hline
    \multicolumn{1}{|c|}{} & \multicolumn{3}{|c|}{data set details} &  \multicolumn{3}{|c|}{runtime} \\
    \hline
    Data set & $m$ & $p$ & $k$ & \multicolumn{1}{|c|}{Prot 1} & \multicolumn{1}{|c|}{Prot 1, Ln 9} & \multicolumn{1}{|c|}{Prot 2} \\
\hline
    CogLoad & 632 & 120 & 12 & 27 sec & 23 sec & 1.13 sec  \\
    \hline
    LSVT & 126 & 310 & 103 & 152 sec & 53 sec & 0.33 sec  \\
    \hline
    SPEED & 8,378 & 122 & 67 & 1,837 sec & 1,812 sec & 14.73 sec  \\
    \hline
    \end{tabular}
    }
\end{table}

\section{Experiments and Results}
\label{sec:exp}
%\blue{We need a table with results that show that feature selection based on Gini mean-split helps to improve the accuracy. Ideally, this table should have a variety of data sets. The results can be computed with sklearn, i.e.~in the clear, as long as the set of selected features are the same in sklearn and in MP-SPDZ. Please add this as soon as you can.}

The first four columns of Table \ref{table:acc_tab} contain details for three data sets corresponding to binary classification tasks with continuous valued input features: Cognitive Load Detection\footnote{\tiny\url{https://www.ubittention.org/2020/data/Cognitive-load\%20challenge\%20description.pdf}} (CogLoad) \cite{Gjoreski_2020}, Lee Silverman Voice Treatment\footnote{\tiny\url{https://archive.ics.uci.edu/ml/datasets/LSVT+Voice+Rehabilitation}} (LSVT) \cite{LSVT}, and Speed Dating\footnote{\tiny\url{https://www.openml.org/d/40536}} (SPEED) \cite{speed_dating}, along with the number of instances $m$, raw features $p$, selected features $k$, and folds for cross-validation (CV).
%are shown in first four colunmns of Table \ref{table:acc_tab} as $m$, $p$, $k$ and fold.
%
The middle five columns of Table \ref{table:acc_tab} contain accuracy results by averaging from CV for logistic regression (LR) models trained on the RAW data sets with all $p$ features, and on reduced data sets with only the top $k$ features selected with a variety of scoring techniques, namely MS-GINI (as proposed in this paper), traditional Gini impurity (GI), Pearson correlation coefficient (PCC), and mutual information (MI). Feature selection with all these techniques was performed according to the filter approach, i.e.~independently of the fact that the selected features were subsequently used to train a LR model. As the results show, feature selection based on MS-GINI is on par with the other methods, and substantially improves the accuracy compared to model training on the RAW data sets.

%We train logistic regression models implemented by Scikit-learn \cite{scikit-learn} on data processed by different methods in plaintext manner. Accuracy results of those models for each data set are shown in last five columns of Table \ref{table:acc_tab} and the best model for each data set is in bold. We can see models on the data processed by filter-based feature selection methods perform better than models on the raw data (RAW) and models based on the MS-GINI method perform better than model on most other methods. {\color{red} What are GI, PCC and MI? This is not explained here nor in the table}

The last three columns of Table \ref{table:acc_tab} contain runtime results for protocol $\PPGINIFS$ for secure filter-based feature selection with MS-GINI
(see Protocol \ref{prtc:msgini_fs}). To obtain these results, we implemented $\PPGINIFS$ along with the supporting protocols $\PPGINI$ and $\PPFILTERFS$ in MP-SPDZ \cite{cryptoeprint:2020:521}. 
 All benchmark tests were completed on 3 or 4 co-located F32s V2 Azure virtual machines. Each VM contains 32 cores, 64 GiB of memory, and up to a 14 Gbps network bandwidth between each virtual machine.
The runtime results are for semi-honest (``passive'') and malicious (``active'') adversary models (see Sec.~\ref{sec:smc}) in a 3PC or 4PC honest-majority setting over a ring $\mathbb{Z}_q$ with $q = 2^{64}$.
%with parties \textit{Alice}, \textit{Bob}, and \textit{Carol} shown in Figure \ref{fig:3pc}. 
Each of the parties ran on separate machines, which means that the results in Table \ref{table:acc_tab} cover communication time in addition to computation time. Similarly as for the accuracies, the reported runtimes in Table \ref{table:acc_tab} are an average across the folds. The relative differences between the passive 3PC, active 3PC , and active 4PC settings are in line with known findings from the MPC literature, in particular the fact that completing private feature selection in the active setting takes substantially longer than in the passive setting; this increase in runtime is a price one has to pay for security and correctness in case the parties can not be trusted to follow the protocol instructions.

For further insight in the dominating factors in the runtime cost, in Table \ref{table:perf_tab} we present more fine-grained runtime results for the active 3PC setting. Protocol 2, which is executed once per feature, in itself grows in the number of instances $m$. While the nested for-loop on Line 1-8 in Protocol 1 depends on $k$ and $p$ only, the matrix multiplication on Line 9 in Protocol 1 depends on all of $m$, $p$, and $k$, and contributes substantially to the runtime.
%As expected, the runtime increases in terms of the number of instances $m$, the number of original features $p$, and the number of selected features $k$. 
The increase in runtime for the SPEED vs.~the CogLoad data set e.g., which have almost the same number of original features $p$, is due both to the increase in $m$ (which affects Line 9 in Protocol 1, and Line 3-11 in Protocol 2), and the increase in $k$ (which affects Line 1-8 of Protocol 1).

\section{Conclusion and Future Work}
\label{sec:conclusion}
Data preprocessing, an important part of the ML model development pipeline, has been largely overlooked in the PPML literature to date. In this paper we have proposed an MPC protocol for privacy-preserving selection of the top $k$ features of a data set, and we have demonstrated its feasibility in practice through an experimental evaluation. Our protocol is based on the filter approach for feature selection, which means that it is independent of any specific ML model architecture. Furthermore, it can be used in combination with any feature scoring technique. In this paper, we have proposed an efficient MPC protocol based on Gini impurity to this end.

In addition to MPC protocols for other feature selection techniques,
MPC protocols for many more tasks related to the data preprocessing phase still need to be developed, including privacy-preserving hyperparameter search to determine the best value of $k$ for the number of features to be selected, as well as protocols for dealing with outliers and missing values. While these may be perceived as less exciting tasks of the ML end-to-end pipeline, they are crucial to enable PPML applications in practical data science.

%\section*{Acknowledgements}
%The authors would like to thank Marcel Keller for making the MP-SPDZ framework available, and for his assistance in the use of the framework.
%
%The authors would like to thank Microsoft for the generous donation of cloud computing credits through the UW Azure Cloud Computing Credits for Research program.

% In the unusual situation where you want a paper to appear in the
% references without citing it in the main text, use \nocite
%\nocite{langley00}

\bibliographystyle{plain}
\bibliography{references}

\begin{thebibliography}{10}

\bibitem{escudero2020}
Mark Abspoel, Daniel Escudero, and Nikolaj Volgushev.
\newblock Secure training of decision trees with continuous attributes.
\newblock In {\em Proceedings on Privacy Enhancing Technologies (PoPETs)},
  pages 167--187, 2021.

\bibitem{agarwal2019protecting}
Anisha Agarwal, Rafael Dowsley, Nicholas~D McKinney, Dongrui Wu, Chin~Teng Lin,
  Martine De~Cock, and Anderson Nascimento.
\newblock Protecting privacy of users in brain-computer interface applications.
\newblock {\em IEEE Transactions on Neural Systems and Rehabilitation
  Engineering}, 27(8):1546--1555, 2019.

\bibitem{agrawal2019quotient}
Nitin Agrawal, Ali Shahin~Shamsabadi, Matt~J Kusner, and Adri{\`a} Gasc{\'o}n.
\newblock {QUOTIENT}: two-party secure neural network training and prediction.
\newblock In {\em ACM SIGSAC Conference on Computer and Communications
  Security}, pages 1231--1247, 2019.

\bibitem{10.1145/2976749.2978331}
Toshinori Araki, Jun Furukawa, Yehuda Lindell, Ariel Nof, and Kazuma Ohara.
\newblock High-throughput semi-honest secure three-party computation with an
  honest majority.
\newblock In {\em Proceedings of the 2016 ACM SIGSAC Conference on Computer and
  Communications Security}, page 805–817, 2016.

\bibitem{banerjee2011privacy}
Madhushri Banerjee and Sumit Chakravarty.
\newblock Privacy preserving feature selection for distributed data using
  virtual dimension.
\newblock In {\em Proceedings of the 20th ACM International Conference on
  Information and Knowledge Management}, pages 2281--2284, 2011.

\bibitem{Bogdanov2013ObliviousSO}
Dan Bogdanov, Sven Laur, and Riivo Talviste.
\newblock Oblivious sorting of secret-shared data.
\newblock {\em Technical Report}, 2013.

\bibitem{breiman1984}
Leo Breiman, Jerome Friedman, Charles Stone, and Richard Olshen.
\newblock {\em Classification and Regression Trees}.
\newblock Taylor and Francis, 1st edition, 1984.

\bibitem{catrina2010improved}
O.~Catrina and S.~De~Hoogh.
\newblock Improved primitives for secure multiparty integer computation.
\newblock In {\em International Conference on Security and Cryptography for
  Networks}, pages 182--199. Springer, 2010.

\bibitem{fixed_point_conversion}
O.~Catrina and A.~Saxena.
\newblock Secure computation with fixed-point numbers.
\newblock In {\em 14th International Conference on Financial Cryptography and
  Data Security}, volume 6052 of {\em Lecture Notes in Computer Science}, pages
  35--50. Springer, 2010.

\bibitem{CHANDRASHEKAR201416}
Girish Chandrashekar and Ferat Sahin.
\newblock A survey on feature selection methods.
\newblock {\em Computers \& Electrical Engineering}, 40(1):16 -- 28, 2014.

\bibitem{Choudhary:2020}
C.A. Choudhary, M.~{De Cock}, R.~Dowsley, A.~Nascimento, and D.~Railsback.
\newblock Secure training of extra trees classifiers over continuous data.
\newblock In {\em AAAI-20 Workshop on Privacy-Preserving Artificial
  Intelligence}, 2020.

\bibitem{mpc_book}
Ronald Cramer, Ivan~Bjerre Damgard, and Jesper~Buus Nielsen.
\newblock {\em Secure Multiparty Computation and Secret Sharing}.
\newblock Cambridge University Press, 1st edition, 2015.

\bibitem{cryptoeprint:2020:1330}
A.~Dalskov, D.~Escudero, and M.~Keller.
\newblock Fantastic four: Honest-majority four-party secure computation with
  malicious security.
\newblock Cryptology ePrint Archive, Report 2020/1330, 2020.

\bibitem{dalskov2019secure}
A.~Dalskov, D.~Escudero, and M.~Keller.
\newblock Secure evaluation of quantized neural networks.
\newblock {\em Proceedings on Privacy Enhancing Technologies},
  2020(4):355--375, 2020.

\bibitem{AISec:CDNN15}
Martine {De Cock}, Rafael Dowsley, Anderson C.~A. Nascimento, and Stacey~C.
  Newman.
\newblock Fast, privacy preserving linear regression over distributed datasets
  based on pre-distributed data.
\newblock In {\em 8th ACM Workshop on Artificial Intelligence and Security
  (AISec)}, page 3–14, 2015.

\bibitem{BMCMG:CDNRST21}
Martine {De Cock}, Rafael Dowsley, Anderson C.~A. Nascimento, Davis Railsback,
  Jianwei Shen, and Ariel Todoki.
\newblock {High performance logistic regression for privacy-preserving genome
  analysis}.
\newblock {\em BMC Medical Genomics}, 14(1):23, 2021.

\bibitem{de2014practical}
Sebastiaan De~Hoogh, Berry Schoenmakers, Ping Chen, and Harm op~den Akker.
\newblock Practical secure decision tree learning in a teletreatment
  application.
\newblock In {\em International Conference on Financial Cryptography and Data
  Security}, pages 179--194. Springer, 2014.

\bibitem{speed_dating}
Raymond Fisman, Sheena~S. Iyengar, Emir Kamenica, and Itamar Simonson.
\newblock Gender differences in mate selection: Evidence from a speed dating
  experiment.
\newblock {\em The Quarterly Journal of Economics}, 121(2):673--697, 2006.

\bibitem{Gjoreski_2020}
Martin Gjoreski, Tine Kolenik, Timotej Knez, Mitja Luštrek, Matjaž Gams,
  Hristijan Gjoreski, and Veljko Pejović.
\newblock Datasets for cognitive load inference using wearable sensors and
  psychological traits.
\newblock {\em Applied Sciences}, 10(11):38--43, 2020.

\bibitem{zigzag}
M.~Goodrich.
\newblock Zig-zag sort: A simple deterministic data-oblivious sorting algorithm
  running in o(n log n) time.
\newblock In {\em Proceedings of the 46th Annual ACM Symposium on Theory of
  Computing}, pages 684--693, 2014.

\bibitem{guo2020secure}
Chuan Guo, Awni Hannun, Brian Knott, Laurens van~der Maaten, Mark Tygert, and
  Ruiyu Zhu.
\newblock Secure multiparty computations in floating-point arithmetic.
\newblock {\em arXiv preprint arXiv:2001.03192}, 2020.

\bibitem{jafer2015framework}
Yasser Jafer, Stan Matwin, and Marina Sokolova.
\newblock A framework for a privacy-aware feature selection evaluation measure.
\newblock In {\em 13th Annual Conference on Privacy, Security and Trust (PST)},
  pages 62--69. IEEE, 2015.

\bibitem{cryptoeprint:2020:521}
Marcel Keller.
\newblock {MP-SPDZ}: A versatile framework for multi-party computation.
\newblock In {\em Proceedings of the 2020 ACM SIGSAC Conference on Computer and
  Communications Security}, page 1575–1590, 2020.

\bibitem{kumar2020cryptflow}
N.~Kumar, M.~Rathee, N.~Chandran, D.~Gupta, A.~Rastogi, and R.~Sharma.
\newblock {CrypTFlow}: Secure {TensorFlow} inference.
\newblock In {\em 41st IEEE Symposium on Security and Privacy}, 2020.

\bibitem{cogLoad}
Xiling Li and Martine {De Cock}.
\newblock Cognitive load detection from wrist-band sensors.
\newblock In {\em Adjunct Proceedings of the 2020 ACM International Joint
  Conference on Pervasive and Ubiquitous Computing and Proceedings of the 2020
  ACM International Symposium on Wearable Computers}, page 456–461, 2020.

\bibitem{lindell2000privacy}
Yehuda Lindell and Benny Pinkas.
\newblock Privacy preserving data mining.
\newblock In {\em Annual International Cryptology Conference}, pages 36--54.
  Springer, 2000.

\bibitem{NYT:2014}
Steven Lohr.
\newblock For big-data scientists, ‘janitor work’ is key hurdle to
  insights.
\newblock The New York Times, 2014.

\bibitem{7958569}
P.~{Mohassel} and Y.~{Zhang}.
\newblock Secureml: A system for scalable privacy-preserving machine learning.
\newblock In {\em IEEE Symposium on Security and Privacy (SP)}, pages 19--38,
  2017.

\bibitem{nikolaenko2013privacy}
Valeria Nikolaenko, Udi Weinsberg, Stratis Ioannidis, Marc Joye, Dan Boneh, and
  Nina Taft.
\newblock Privacy-preserving ridge regression on hundreds of millions of
  records.
\newblock In {\em IEEE Symposium on Security and Privacy (SP)}, pages 334--348,
  2013.

\bibitem{Rao2019SecureTF}
Vanishree Rao, Yunhui Long, Hoda Eldardiry, Shantanu Rane, Ryan~A. Rossi, and
  Frank Torres.
\newblock Secure two-party feature selection.
\newblock {\em arXiv preprint arXiv:1901.00832}, 2019.

\bibitem{riazi2018chameleon}
M.S. Riazi, C.~Weinert, O.~Tkachenko, E.M. Songhori, T.~Schneider, and
  F.~Koushanfar.
\newblock Chameleon: A hybrid secure computation framework for machine learning
  applications.
\newblock In {\em Asia Conference on Computer and Communications Security},
  pages 707--721, 2018.

\bibitem{sheikhalishahi2017privacy}
Mina Sheikhalishahi and Fabio Martinelli.
\newblock Privacy-utility feature selection as a privacy mechanism in
  collaborative data classification.
\newblock In {\em IEEE 26th International Conference on Enabling Technologies:
  Infrastructure for Collaborative Enterprises (WETICE)}, pages 244--249, 2017.

\bibitem{LSVT}
Athanasios Tsanas, {Max A.} Little, Cynthia Fox, and {Lorraine O.} Ramig.
\newblock Objective automatic assessment of rehabilitative speech treatment in
  parkinson's disease.
\newblock {\em IEEE Transactions on Neural Systems and Rehabilitation
  Engineering}, 22(1):181--190, 2014.

\bibitem{wagh2019securenn}
Sameer Wagh, Divya Gupta, and Nishanth Chandran.
\newblock {SecureNN}: 3-party secure computation for neural network training.
\newblock {\em Proceedings on Privacy Enhancing Technologies (PoPETs)},
  2019(3):26--49, 2019.

\bibitem{ye2019distributed}
Xiucai Ye, Hongmin Li, Akira Imakura, and Tetsuya Sakurai.
\newblock Distributed collaborative feature selection based on intermediate
  representation.
\newblock In {\em International Joint Conference on Artificial Intelligence},
  pages 4142--4149, 2019.

\end{thebibliography}

\end{document}